\documentclass[conference]{IEEEtran}
\IEEEoverridecommandlockouts
\usepackage{cite}
\usepackage{amsmath, amssymb, amsfonts, mathrsfs}
\usepackage{algorithmic}
\usepackage{times}
\usepackage[margin=1in]{geometry}
\usepackage[normalem]{ulem} 
\usepackage{graphicx, color}
\usepackage{textcomp}
\usepackage{xcolor}
\usepackage{multirow}

\newcommand{\eat}[1]{}

\begin{document}

\title{\huge{Designing Machine Learning Surrogates using Outputs of Molecular Dynamics Simulations as Soft Labels}\\}

\author{\IEEEauthorblockN{
JCS Kadupitiya,
Nasim Anousheh,
Vikram Jadhao
}
\IEEEauthorblockA{\textit{Intelligent Systems Engineering} \\
\textit{Indiana University}\\
Bloomington, Indiana 47408 \\
\{kadu,nanoushe,vjadhao\}@iu.edu}
}

\maketitle

\begin{abstract}
Molecular dynamics simulations are powerful tools to extract the microscopic mechanisms characterizing the properties of soft materials.
We recently introduced machine learning surrogates for molecular dynamics simulations of soft materials and demonstrated that artificial neural network based regression models can successfully predict the relationships between the input material attributes and the simulation outputs.
Here, we show that statistical uncertainties associated with the outputs of molecular dynamics simulations can be utilized to train artificial neural networks and design machine learning surrogates with higher accuracy and generalizability. 
We design soft labels for the simulation outputs by incorporating the uncertainties in the estimated average output quantities, and introduce a modified loss function that leverages these soft labels during training to significantly reduce the surrogate prediction error for input systems in the unseen test data.
The approach is illustrated with the design of a surrogate for molecular dynamics simulations of confined electrolytes to predict the complex relationship between the input electrolyte attributes and the output ionic structure. 
The surrogate predictions for the ionic density profiles show excellent agreement with the ground truth results produced using molecular dynamics simulations. 
The high accuracy and small inference times associated with the surrogate predictions provide quick access to quantities derived using the number density profiles and facilitate rapid sensitivity analysis.
\end{abstract}

\begin{IEEEkeywords}
Machine Learning, Molecular Dynamics Simulations, Soft Materials, Deep Learning, Soft Labels
\end{IEEEkeywords}


\section{Introduction}

Molecular dynamics (MD) simulations are powerful computational methods for investigating the microscopic origins of the macroscopic behavior of soft materials such as viral capsids, lipid bilayers, confined electrolytes, self-assembled colloids, and polymers \cite{glotzer2015assembly,perilla2016all,brunk2019computational,jadhao2019rheological}. These simulations furnish molecular-level mechanisms that drive structure and property control in soft materials while isolating interesting regions of the material design space to aid experimental exploration and discovery. 
Despite the use of parallel computing techniques, simulations of soft materials often incur high computational costs. 
Recent years have seen a surge in the integration of machine learning (ML) methods with simulations to reduce their computational costs, enhance their predictive power, and expedite the analysis of high-dimensional output data \cite{ferguson2017machine,glotzer2017,schoenholz2017relationship,guo2018adaptive,wessels2021machine,wang2019machine,casalino2021ai,gbell1,kadupitiya2021probing,moradzadeh2019molecular,aspuru2019,sun2019deep,kasim2020building,kadupitiya2020machine,kadupitiya2020simulating}. ML has been used to develop efficient force fields \cite{wang2019machine,casalino2021ai,gbell1}, reduce high-dimensional simulation data to isolate molecular-level mechanisms  \cite{schoenholz2017relationship,kadupitiya2021probing}, and develop surrogates to accurately predict simulation outcomes and expedite the exploration of the material design space  \cite{moradzadeh2019molecular,aspuru2019,sun2019deep,kasim2020building}.
The last set of applications, where surrogates are designed for learning the relationship between the input variables and simulation outputs, is the subject of this paper. 

Machine learning has been used to design surrogates to predict outputs of a wide variety of simulations \cite{moradzadeh2019molecular,aspuru2019,sun2019deep,kasim2020building,degiacomi2019coupling,rahman2021machine}. Deep neural networks trained on simulation data accurately predicted adsorption equilibria for different thermodynamic conditions \cite{sun2019deep}. ML surrogate for \emph{ab initio} MD simulations bypassed the explicit time evolution of the particle trajectories to predict the dissociation timescale of compounds \cite{aspuru2019}. Convolutional neural network based ML ``emulators'' have been developed to predict outputs of simulations in different scientific domains including climate science and biogeochemistry \cite{kasim2020building}. Deep neural network-based denoising autoencoder was trained to successfully predict the temporally-averaged radial distribution function of Lennard-Jones fluids from a single snapshot of the radial distribution function generated in MD simulations \cite{moradzadeh2019molecular}. 
A convolutional neural network-based surrogate model was trained on data produced by MD simulation of polymer materials to predict filled rubber material properties that take the kinetic information of filler morphology as input \cite{kojima2020synthesis}.

We recently introduced ML surrogates for MD simulations of soft materials \cite{kadupitiya2019machine,kadupitiya2020machine2,fox2019learning,jadhao2020integrating}. 
We demonstrated that artificial neural network based regression models trained on data from completed MD simulations of soft-matter systems can successfully predict the relationships between the input parameters and the simulation outcomes, bypassing part or all of the explicit evolution of the simulated components. 
The approach was illustrated using simulations of electrolyte solutions confined by material surfaces. 
A key goal of the computational studies of confined electrolytes is to establish the links between the output distributions of electrolyte ions and the input electrolyte attributes for diverse solution conditions.
These links provide a reliable guide to the regions of the material design space where significant changes in the structural organization of ions are expected, which can aid the experimental exploration and design of electrolyte-based materials \cite{he2009tuning,faucher2019critical,park2017maximizing,werber2016materials,levin1,jos4}.  
The ML surrogate was trained to learn the relationship between the output distribution of positive ions and input variables characterizing the electrolytes containing positive and negative ions of the same size confined by uncharged surfaces. 
The ionic density profiles predicted by the surrogate were found to be in excellent agreement with MD simulations  \cite{kadupitiya2019machine,kadupitiya2020machine2}. Further, the inference time associated with the predictions was $10000\times$ less than the corresponding MD simulation runtime. 

In this paper, we extend the design and application of ML surrogates for MD simulations of soft materials in many significant ways.
We show that statistical uncertainties present in the outputs of MD simulations can be utilized to train deep neural networks and design ML surrogates with higher accuracy and generalizability.
We design soft labels for the simulation outputs by incorporating the uncertainties in the estimated average output quantities, and introduce a modified loss function that 
leverages these soft labels during training to significantly reduce the surrogate prediction error for input systems in the unseen test data. 
Using such soft labels for the ground truth data bypasses the need to explicitly expand the training dataset size in order to generalize the surrogate to learn complex relationships between input soft material attributes and output structural quantities.

The approach is illustrated with the design of a surrogate for MD simulations of confined electrolytes that predicts the distribution of positive and negative ions in confinement. The complexity of the relationship between input electrolyte attributes and output ionic structure is significantly enhanced by including ionic size asymmetry, charged surfaces, and a broad range of concentrations in the electrolyte model \cite{anousheh2020ionic} (the surrogate developed in our previous paper \cite{kadupitiya2020machine2} only predicted the distribution of positive ions for electrolytes with size-symmetric ions confined by uncharged interfaces).
We show that the surrogate, trained using the modified loss function that leverages the statistical uncertainties in the output ionic distributions, predicts the number densities of positive and negative ions in excellent agreement with the ground truth MD simulation results.

The high accuracy and small inference times (a factor of $200,000\times$ smaller than the corresponding simulation time) associated with the surrogate predictions of number density profiles enable quick access to structural quantities such as the net charge density and the integrated charge \cite{qiao2004charge,dopke2019preferential}. 
The surrogate-enabled generation of output ionic structure for millions of input electrolyte systems within minutes facilitates the implementation of complex analysis tasks such as sensitivity analysis that help distill the contributions of the input ionic attributes to the output ionic structure.

The rest of this paper is organized as follows. 
Section \ref{sec:ml.surrogate} describes the details of designing the ML surrogate. Section \ref{sec:results} discusses the accuracy characteristics and generalizability of the ML surrogate and presents the results of its application in confined electrolytes. Section \ref{sec:conclusion} presents conclusions. 

\begin{figure}
\centerline{\includegraphics[scale=0.38]{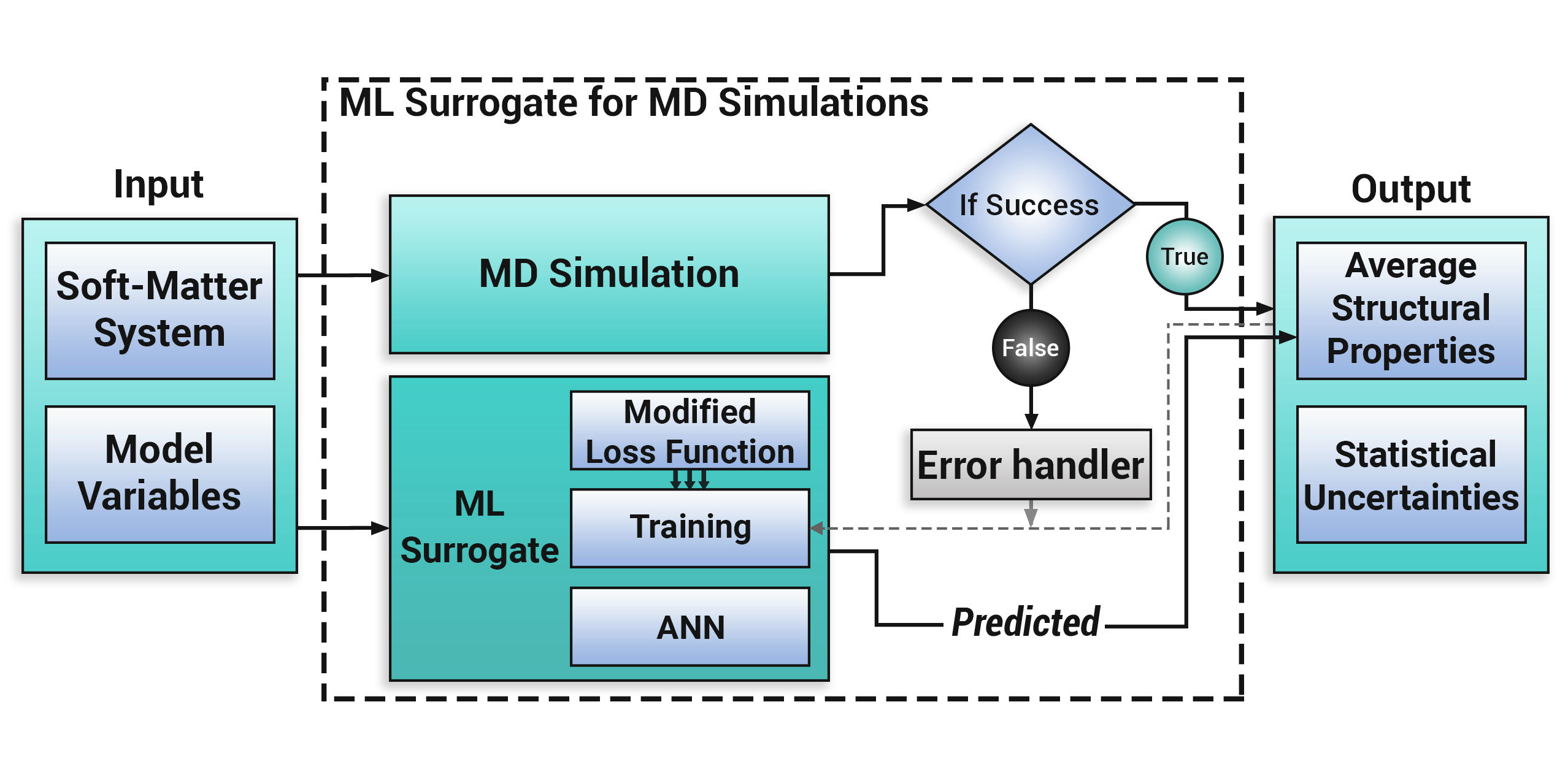}}
\caption
{\label{fig.overview}
System overview of the machine learning (ML) surrogate for molecular dynamics (MD) simulations approach to generate predictions for the structural properties (output) of soft-matter systems (input).
}
\end{figure}

\section{Machine Learning Surrogates for Molecular Dynamics Simulations}\label{sec:ml.surrogate}

Figure \ref{fig.overview} shows a system overview of the approach to design and use ML surrogates for MD simulations of soft materials \cite{kadupitiya2019machine,kadupitiya2020machine2}.
We first run MD simulations for different model variables (input) characterizing the soft-matter system, and save the desired simulation outcomes (output) for training the ML surrogate, which occurs after a set number of successful simulation runs. 
Once trained, the ML surrogate is used to predict the relationship between the input variables and the output structural properties.
Figure \ref{fig.overview} highlights the modified loss function introduced in this paper to train the surrogate by leveraging the statistical uncertainties present in the MD simulation output data.

In the specific example of designing the surrogate for MD simulations of confined electrolytes, the input variables are confinement length, electrolyte concentration, ionic sizes, and surface charge, and the structural properties are the ionic density profiles. 
We begin by briefly describing these input variables and output quantities.
A more detailed description of the model variables and the simulation methods and outputs can be found in our previous papers  \cite{anousheh2020ionic,jing2015ionic}. 

\subsection{Input Variables and Output Quantities}\label{sec:system.inputs.outputs}

We consider a monovalent electrolyte in water confined by two planar interfaces at room temperature $T=298$ K.
The planes at $z = -h/2$ and $z = h/2$ represent the location of the interfaces with $h$ being the interfacial separation ($z=0$ corresponds to the midpoint between the interfaces).
Each interface is characterized with a uniform charge density $\sigma_s < 0$. A coarse-grained model \cite{boda,allen1,tyagi,luijten.jcp2014,jing2015ionic} is employed to describe the electrolyte solution. Water is modeled as an implicit solvent with a relative dielectric permittivity of $80$. The positively-charged ions (cations) and negatively-charged ions (anions) associated with the monovalent electrolyte are modeled as spheres with hydrated sizes $d_+$ and $d_-$ respectively. An appropriate number of counterions, modeled as cations of the same diameter and charge,  are included in the confinement to ensure electroneutrality. 

Five input parameters: $h$, electrolyte concentration $c$, $d_+$, $d_-$, and $\sigma_s$, are used to design and train the ML surrogate (Table \ref{tab:input.params}).
$c$ is defined as $c = N_- / V$, where $N_-$ is the number of anions and $V$ is the volume of the simulation box. 
We note that not all ionic attributes and solutions conditions that are expected to alter the output ionic structure are considered as tunable input variables. 
For example, ion valencies (set to $1, -1$), temperature (set at 298 K) and solvent permittivity (set to 80) are held fixed across all the simulations. 

All MD simulations are performed using LAMMPS \cite{lammps.plimpton} in an NVT ensemble at $T=298$ K. Ion-ion and ion-interface steric interactions are modeled using Lennard-Jones potentials \cite{jing2015ionic}, and all electrostatic interactions are modeled using Coulomb potentials whose long-range is properly treated with Ewald sums \cite{deserno1998mesh}. 
Each system is simulated for 1 ns to reach equilibrium with a timestep of $1$ femtosecond. After equilibration, systems are simulated for at least $\approx 9$ ns, and trajectory data are collected every $0.1$ ps. 

Samples from the trajectory data are used to compute the average number densities $n_{+}(z)$ and $n_{-}(z)$ of cations and anions, respectively, together with the associated statistical uncertainties $\epsilon_{+}(z)$ and $\epsilon_{-}(z)$. 
Note that due to the planar symmetry and the homogeneously-charged interfaces, the ionic distributions vary only in the direction perpendicular to the interfaces, and hence are functions of a single variable $z$. $n_{+}(z)$ of cations and $n_{-}(z)$ of anions form the output of the ML surrogate. 
These number densities are used to derive quantities such as the net charge density $\rho(z)$:
\begin{equation}\label{eq:rho}
\rho(z) = en_{+}(z) - en_{-}(z),
\end{equation}
and the integrated charge $S(z)$: 
\begin{equation}\label{eq:S}
S(z) = \sigma_{s} + \int_{-h/2}^{z} \rho(z') dz'.
\end{equation}
Note that $S(z)$ is defined using the left planar interface as the reference surface and is meaningful for $-h/2 \le z \le 0$, i.e., for $z$ up to the center of the confined region \cite{anousheh2020ionic}. 

\begin{table}
\caption{Input variables and associated ranges.}
\begin{tabular*}{0.48\textwidth}{@{\extracolsep{\fill}}llll}
    \hline
    Input Variable & Range \\
    \hline
    Interfacial separation ($h$)   &  4 -- 5  nm \\
    Electrolyte concentration ($c$)   & 0.1 -- 2.0 M \\
    Cation diameter ($d_+$)     & 0.2 -- 0.63 nm \\
    Anion diameter ($d_-$)     & 0.2 -- 0.63 nm \\
    Surface charge density ($\sigma_{s}$) & $-$0.01, $-$0.015, $-$0.02 $\mathrm{C}/\mathrm{m}^2$ \\
    \hline
  \end{tabular*}
\label{tab:input.params}
\end{table}

\subsection{Data Generation and Preprocessing} \label{sec:dataprep}
Table \ref{tab:input.params} shows the ranges over which the input parameters $h$, $c$, $d_+$, $d_-$, and $\sigma_s$ are tuned to generate the dataset for training and testing the ML surrogate. 
By sweeping over a few discrete values for each input variable within the associated range, 4,050 unique input configurations are created. For each of these configurations, MD simulations are run and the converged distributions of positive and negative ions are extracted as output using the ion trajectory samples collected during the MD simulation.
These distributions are specified by computing the average population densities of ions at $502$ points (bins) within the confinement region (length) $z \in (-h/2, h/2)$. The trajectory samples are also used to compute the associated statistical uncertainties in the extracted average number densities of cations and anions.
Each MD simulation is performed for over $\approx 10$ nanoseconds and takes 
$\approx 6$ hours using LAMMPS with 96 cores. Generating the entire dataset (size $\approx 15$ GB) took approximately 20 days, including the queue wait times on the Indiana University BigRed3 supercomputing cluster. 

The ML surrogate is trained to predict the average number densities of cations and anions for all $502$ bins. Thus, the surrogate makes a total of $P = 1004$ predictions to quantify the output ionic distributions. The entire dataset is separated into training and testing sets using a ratio of $0.8$ $:$ $0.2$, resulting in training and testing datasets of 3240 and 810 configurations, respectively. A min-max normalization filter is applied to normalize the input data at the preprocessing stage.

\begin{figure}
\centerline{\includegraphics[scale=0.54]{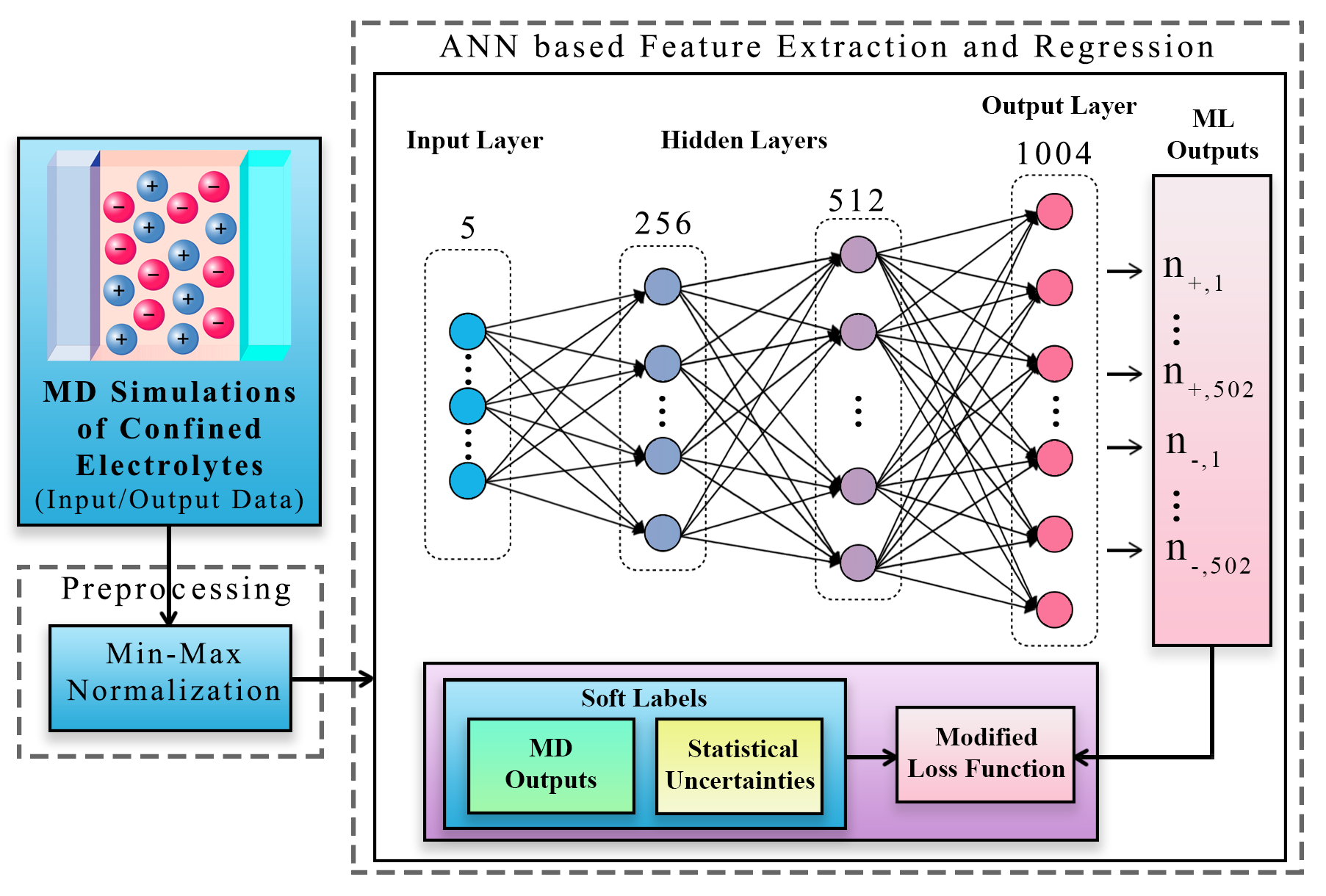}}
\caption
{\label{ml.overview}
Artificial neural network (ANN) based regression model used in the machine learning (ML) surrogate to predict output density profiles of cations ($n_+$) and anions ($n_-$). ANN is trained using the modified loss function that treats the molecular dynamics (MD) simulation outputs as soft labels and leverages the associated uncertainties to determine the optimal weights and biases of the hidden layers.
}
\end{figure}

\subsection{Feature Extraction and Regression}
Figure \ref{ml.overview} shows the artificial neural network (ANN) architecture employed in the ML surrogate to implement the regression and prediction of the desired $P = 1004$ output quantities. The ANN has 2 hidden layers, similar to the surrogate designed in our earlier work \cite{kadupitiya2020machine2}.   
The regression process determines the weights and biases in these hidden layers following an error backpropagation algorithm, implemented via a stochastic gradient descent procedure.
This process uses a training dataset and an appropriate loss function for error computation and backpropagation to update the weights and biases after each batch of input data is regressed through the network by a simple forward prediction.

A common choice for the loss function in regression applications is the mean square error (MSE) between the ground truth data $y_i$ and the predictions $\hat{y}_i$: 
\begin{equation}\label{eq.mse}
L = \frac{1}{b}\sum_{i = 1}^{b} \left( y_i - \hat{y}_i \right)^2,
\end{equation}
where $i$ denotes the input configuration and $b$ is the batch size. Note that $y_i$ and $\hat{y}_i$ are vectors associated with input system $i$ whose size is equal to the number of output dimensions. In our earlier paper, we employed $L$ as the loss function \cite{kadupitiya2020machine2} to train the ANN model to learn the density profiles of positive ions for electrolyte systems involving size-symmetric ions and uncharged interfaces. 
In the present study, the ANN model is trained to learn a more complex input-output relationship due to the consideration of predicting both cation and anion density profiles for electrolytes involving size-asymmetric ions, charged interfaces, and a wider range of concentrations. This rise in complexity is reflected in the larger errors in surrogate predictions for the test data when the ANN model is trained using $L$ (see Section \ref{sec:evaluation}). These errors can be reduced by increasing the number of training data samples, which requires running more MD simulations. 
Here, we adopt a different approach that bypasses the need to run these additional simulations.
We develop a modified loss function that leverages the statistical uncertainties in the ground truth data to reduce the prediction errors and increase the generalizability of the ANN model.

\subsubsection{Modified Loss Function}
Many simulation outcomes such as density profiles of ions are associated with statistical uncertainties. 
These uncertainties are typically extracted as the standard error (standard deviation from the mean) using the samples (e.g., ion positions) collected during the simulation.
For example, for a given input configuration, we extract the average (mean) number density of cations, $n_+(z)$, and the associated statistical uncertainties $\epsilon_{+}(z)$, where the latter are given as:
\begin{equation}\label{eq:error}
    \epsilon_{+}(z) = \left(\frac{1}{S} \sum_{s=1}^{S} \left( n_+^s\left(z\right) - n_+\left(z\right) \right)^2\right)^{1/2}.
\end{equation}
Here, $n_+^s(z)$ is the cation distribution computed using the trajectory sample $s$, and $S$ is the number of samples. Trajectory samples are a set of ion positions collected every $f$ simulation time steps. Generally, $\epsilon_{+}(z)$ includes errors resulting from the time discretization and the errors associated with the sampling process.
The standard error $\epsilon_{+}(z)$ captures the expected variations in the mean $n_+(z)$. 
By only considering the mean $n_+(z)$ as the ground truth data while training a deep learning model, standard loss functions such as $L$ do not leverage the information encoded in $\epsilon_{+}(z)$ regarding the variations in the ground truth data. 
    
Our key idea is to utilize the uncertainties associated with the ground truth data to ``informate'' the training of the ANN model in order to enhance its generalizability when tested on unseen input samples. 
To implement the idea, we modify the standard MSE loss function $L$ by replacing the ground truth data vector $y_i$ with an associated set of normal distributions having mean vector $y_i$ and variance vector $\epsilon_i^2$, where $\epsilon_i$ is the uncertainty vector corresponding to $y_i$. In other words, we replace the ``hard'' labels $y_i$ with \emph{soft labels} $Y_i$, which are randomly sampled from the aforementioned normal distributions.
For example, for each value of $z$, we replace $n_+(z)$ with samples drawn from a normal distribution $\mathcal{N}$ having mean $n_+(z)$ and variance  $\epsilon_{+}^2(z)$. 
In general, the modified loss function $\mathcal{L}$ can be written as:
\begin{equation}\label{eq.mse.modified}
\mathcal{L} = \frac{1}{b}\sum_{i = 1}^{b} \left(Y_i - \hat{y}_i \right)^2, 
\end{equation}
where $Y_i \sim \mathcal{N}(y_i, \epsilon_i^2)$ is a random variable sampled from the normal distribution $\mathcal{N}$.
The back-propagated error through the ANN model obtained as the partial derivative of  $\mathcal{L}$ with respect to the model output $\hat{y}$ is
\begin{equation}\label{eq.mse.partial}
-\frac{\partial\mathcal{L}}{\partial \hat{y}_i} \sim   \mathcal{N}(y_i, \epsilon_i^2) - \hat{y_i}.
\end{equation}
Thus, $\mathcal{L}$ enables the consideration of standard errors associated with the mean ground truth data to update the weights and biases when back-propagating through the network. 

The use of soft labels for describing the ground truth data facilitates a sampling mechanism driven by the normal distributions that implicitly expands the dataset with more samples as the model undergoes training over many epochs. 
In other words, we enable the proliferation of samples without explicitly expanding the training dataset.
Here onward, we refer to the model trained with the modified loss function $\mathcal{L}$ as ModNN and a baseline model trained with $L$ as BaseNN.

We note that research in the applications of ML in computer vision and speech recognition has shown that uncertainties in the ground truth labels can be harnessed toward designing improved ML models, in particular, for classification tasks \cite{smyth1994knowledge,aung2018harnessing,kumano2015analyzing,meng2011multi,scherer2013investigating}.
For example, ground truth labels were represented as soft labels using a probability distribution to train neural networks \cite{smyth1994knowledge} and Bayesian networks \cite{kumano2015analyzing} for image classification tasks. Also, support vector machines for classifying voice quality samples yielded higher accuracy when trained on \emph{fuzzy} output (soft labels) compared to standard output \cite{scherer2013investigating}.

Aung et. al. studied the comparison between soft labels and hard labels for training recognition models as classifiers and regressors for computer vision applications \cite{aung2018harnessing}. Their findings showed that soft labels can improve the accuracy for both classification and regression tasks due to the enhanced regularization effect.
Other studies have explored the connection between improved regularization capabilities via the use of soft labels and dropout regularization \cite{wager2013dropout}. The impact of noise in input or output data on the generalization of ML models has also been studied \cite{krogh1992generalization,crammer2009adaptive}.
Our approach introduces an alternate method based on the modification of the loss function to utilize soft labels during the training of deep neural networks for regression tasks. 

\subsubsection{Surrogate Training}

The implementation, training, and testing of the artificial neural network (ANN) model were programmed using scikit-learn, Keras, and TensorFlow ML libraries \cite{chollet2015keras,buitinck2013api,abadi2016tensorflow}. Scikit-learn was used for grid search and feature scaling, Keras was used to save and load models, and TensorFlow was used to define and train the neural network. 
Training the ANN model involves an appropriate selection of hyperparameters such as the number of first hidden layer units $N_1$, the number of second hidden layer units $N_2$, batch size $b$, and the number of epochs $N_e$. 
$b$ is a hyperparameter of the stochastic gradient descent algorithm that controls the number of training samples allowed to pass through the ANN before updating its trainable parameters (e.g., weights, biases).
$N_e$ is a hyperparameter that controls the number of complete passes made through the entire training dataset.

By performing a grid search, we observed that the hyperparameters are optimized to $N_1 = 256$, $N_2 = 512$, $b=32$, and $N_e = 9500$ for both BaseNN and ModNN models. 
The Adam optimizer was used to optimize the error backpropagation process. Using a trial-and-error process, the learning rate of the Adam optimizer and the dropout rate in the dropout layer were set to $0.0002$ and $0.2$, respectively. 
The weights in the hidden layers and the output layer were initialized for better convergence using a Xavier normal distribution at the beginning. This distribution is characterized with 0 mean and $\sigma = 1/(\sqrt{h_i + h_o})$ variance, where $h_i$ and $h_o$ are the input and output sizes of the hidden layers, respectively \cite{glorot2010understanding}. 
Prototype implementation
written using Python/C++ is available on GitHub \cite{github.nanoconfinement}, and a software application utilizing the surrogate is deployed in nanoHUB \cite{kadupitiya2017}.

\section{Results and Discussion}\label{sec:results}

\subsection{Surrogate Accuracy: BaseNN vs ModNN Models}\label{sec:evaluation}
We begin by comparing the generalizability of the ML surrogate trained using the modified loss function $\mathcal{L}$ (ModNN model) vs the surrogate trained using the standard loss function $L$ (BaseNN model).
The generalization ability can be discussed in terms of overfitting characteristics and accuracy assessed by examining the error incurred on surrogate predictions for inputs in the unseen test data.

We first compute the test loss for each epoch as the average mean square error (MSE) incurred in making $P=1004$ predictions describing the cation and anion density profiles associated with $N_{\mathrm{test}} = 810$ input systems in the test data:
\begin{equation}\label{eq.testloss}
\mathrm{MSE} = \frac{1}{N_{\mathrm{test}} P} \sum_{j = 1}^{N_{\mathrm{test}}} \sum_{k=1}^{P} \left( y_{j}^{k} - \hat{y}_{j}^{k} \right)^2.
\end{equation}
Here $\hat{y}_{j}$ is the prediction vector for the input system $j$ comprising $P$ elements, and $y_{j}$ is the corresponding ground truth vector. Note that $\hat{y}_{j}^{k}$ is the $k^{\mathrm{th}}$ prediction of the ion number density for the input system $j$.

\begin{figure}
\centerline{\includegraphics[scale=0.45]{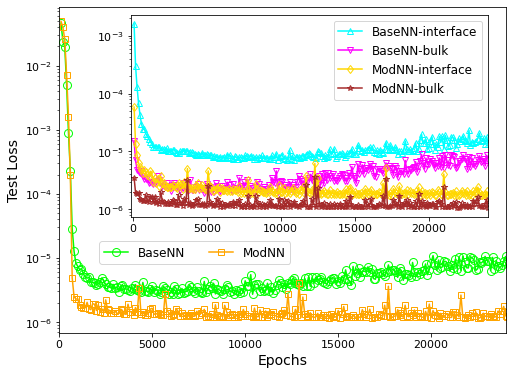}}
\caption
{\label{ml.train.test.loss} Test loss defined in Equation \ref{eq.testloss} vs epochs for BaseNN (green circles) and ModNN (orange squares) models.
MSE of ModNN model is smaller than the BaseNN model for almost all epochs. 
Inset breaks down the test loss into contributions emerging from predictions for the bins near interfaces (bin number $k \in [10,30]$ and $k \in [470,490]$) and those in the bulk near the central region ($k \in [230,270]$). 
Cyan upright triangles and magenta down triangles represent the test loss for the BaseNN model predictions near the interface and in the bulk respectively. Yellow diamonds and brown stars represent the test loss for the ModNN model predictions near the interface and in the bulk respectively.
}
\end{figure}

Figure \ref{ml.train.test.loss} compares the test loss (MSE) for the ModNN vs the BaseNN models. MSE of ModNN model is smaller than the BaseNN model for almost all epochs. 
The test loss decreases to an optimal value of $2.71 \times 10^{-6}$ near 10000 epochs for the BaseNN model, but goes down further to $1.15 \times 10^{-6}$ within 10000 epochs for the ModNN model. 
Within the same number of epochs, we find the optimal training loss for the BaseNN and ModNN models to be $2.83 \times 10^{-6}$ and $1.17 \times 10^{-6}$ respectively. Similar levels of test and train loss values suggest that both models are not overfitted near 10000 epochs \cite{cawley2010over}.
The test loss for the BaseNN model increases gradually when the model is trained beyond 10000 epochs, signaling the crossover to the overfitting territory. In contrast, the test loss for the ModNN model does not cross over to overfitting territory even after 20000 epochs of training.

Figure \ref{ml.train.test.loss} (inset) breaks down the test loss into contributions emerging from predictions near the interfaces (within $\approx 0.2$ nm from either interface) and the predictions in the bulk near the center of the confinement ($-0.1 \lesssim z \lesssim 0.1$ nm). 
The optimal test loss for predictions near the interfaces is higher than that for predictions in the bulk for both BaseNN and ModNN models. 
This difference stems from the relatively higher complexity of the relationship between the interfacial ionic structure and the electrolyte attributes. 

MSE for BaseNN model predictions near the interfaces and in the bulk decreases initially with increasing number of epochs, reaching lowest values of $7.3 \times 10^{-6}$ (interfaces) and $4\times10^{-6}$ (bulk) around 10000 epochs, and then increases gradually on further training.
MSE for ModNN model predictions near the interfaces and in the bulk decreases with increasing number of epochs, reaching $1.7 \times 10^{-6}$ (interfaces) and $1\times10^{-6}$ (bulk) around 10000 epochs, and does not increase on further training.
Compared to the BaseNN model, the ModNN model reduces the prediction error near the interfaces and in bulk by over $6\times$ and over $4\times$ respectively. 

\begin{figure}
    \centering
    \includegraphics[scale=0.45]{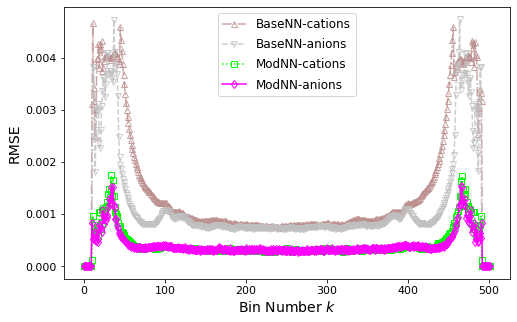}
    \caption{RMSE values extracted using Equation \ref{eq.rmsek} for the predictions made by the BaseNN and ModNN models for cation ($\mathrm{RMSE}_k^{+}$) and anion ($\mathrm{RMSE}_k^{-}$) densities in the $k^{\textrm{th}}$ bin. 
    Brown up triangles and silver down triangles represent the errors $\mathrm{RMSE}_k^{+}$ and $\mathrm{RMSE}_k^{-}$ obtained using the BaseNN model respectively.
    Green squares and magenta diamonds represent $\mathrm{RMSE}_k^{+}$ and $\mathrm{RMSE}_k^{-}$ obtained using the ModNN model respectively. For all $k$, $\mathrm{RMSE}_k^{+}$ and $\mathrm{RMSE}_k^{-}$ produced by the ModNN model are smaller than those produced by the BaseNN model.}
    \label{fig.success}
\end{figure}

We now examine the accuracy associated with the optimal BaseNN and ModNN models produced by selecting weights and biases checkpointed at 9670 epochs using early stopping criteria.
The accuracy is measured for each of the $P = 1004$ ion density predictions associated with an input system by computing the root mean square errors (RMSE). For the $k^{\mathrm{th}}$ prediction, $\mathrm{RMSE}_k$ is the error associated with the prediction of the ion density at the $k^{\mathrm{th}}$ bin:
\begin{equation}\label{eq.rmsek}
\mathrm{RMSE}_k = \left( \frac{1}{N_{\mathrm{test}}} \sum_{j=1}^{N_{\mathrm{test}}} \left(y_j^k - \hat{y}_j^k \right)^2 \right)^{1/2},
\end{equation}
where, $\hat{y}_{j}^{k}$ is the prediction (inference) of the ion number density at the $k^{\mathrm{th}}$ bin for the input system specified by the index $j$, $y_{j}^{k}$ is the corresponding ground truth, and $k = 1, 2, \ldots,$ $1004$. 
$\mathrm{RMSE}_k$ is evaluated by averaging over the errors associated with the $k^{\mathrm{th}}$ bin for all the $N_{\mathrm{test}}$ samples in the test dataset. It is useful to note that the $\mathrm{RMSE}_k$ values are related to the MSE extracted using Equation \ref{eq.testloss} at 9670 epoch via the equality: $\mathrm{MSE} = (1/P)\sum_{k=1}^{P} \mathrm{RMSE}_k^{2}$.

\begin{figure*}
\centerline{\includegraphics[scale=0.5]{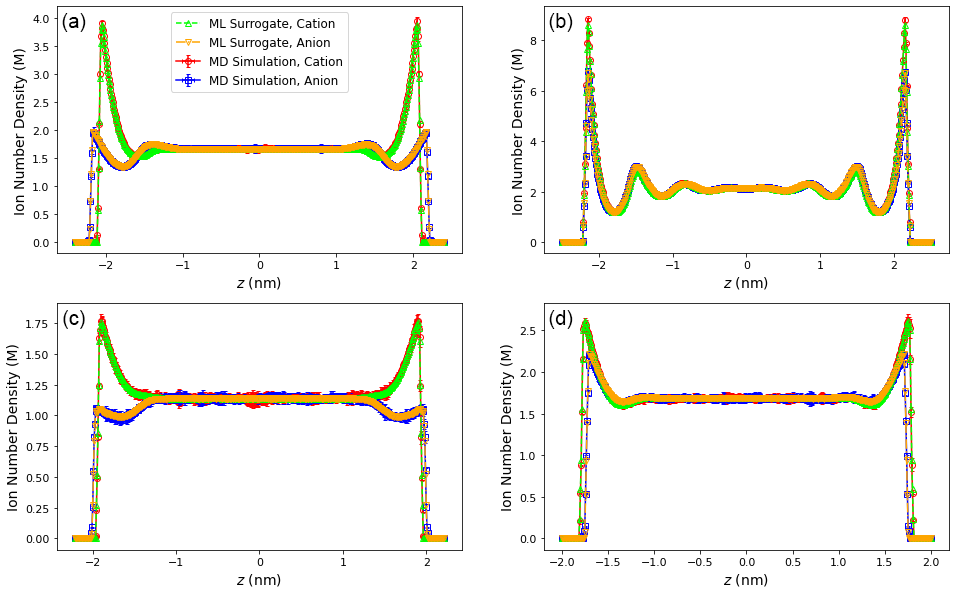}}
\caption
{\label{fig.predictions.density}
Cation and anion density profiles for input electrolyte systems I (a), II (b), III (c), and IV (d) predicted by the ML surrogate vs the ground truth results extracted using MD simulations.
Green up triangles and orange down triangles, respectively, represent the cation and anion densities predicted by the surrogate. Red circles and blue squares with errorbars, respectively, represent the cation and anion densities (with their associated uncertainties) produced by the simulations. Surrogate predictions are in excellent agreement with the ground truth simulation results.
See main text for the input system specifications.
}
\end{figure*}

For each bin, 2 inferences are made: one for the cation density and the other for the anion density, such that half of the predictions ($P/2 = 502$) describe the cation density profile, while the other half describes the anion density profile. We collect these 2 groups of inferences into 2 separate RMSE sets, $\mathrm{RMSE}_k^{+}$ and $\mathrm{RMSE}_k^{-}$, with $k = 1, 2, \ldots, 502$. 
The $k^{\mathrm{th}}$ bin corresponds to the discretization of the confinement length and is correlated with the distance from the left interface. Thus, $\mathrm{RMSE}_k^{+}$ and $\mathrm{RMSE}_k^{-}$ also indicate the variation of the error in the prediction of cation and anion densities as a function of the location within the confinement. 

Figure \ref{fig.success} shows $\mathrm{RMSE}_k^{+}$ and $\mathrm{RMSE}_k^{-}$ associated with the predictions made by the ModNN and BaseNN models for the cation and anion densities in the $k^{\textrm{th}}$ bin. For all $k$, $\mathrm{RMSE}_k^{+}$ and $\mathrm{RMSE}_k^{-}$ values produced by the ModNN model are smaller than those produced by the BaseNN model.
The average $\mathrm{RMSE}^+ = (2/P)\sum_{k=1}^{P/2} \mathrm{RMSE}_k^+$ is 0.0015 and 0.00042 for the BaseNN and ModNN models respectively.
The average $\mathrm{RMSE}^- = (2/P)\sum_{k=1}^{P/2} \mathrm{RMSE}_k^-$ is 0.0012 and 0.00041 for the BaseNN and ModNN models respectively.
$\mathrm{RMSE}_k^{+}$ and $\mathrm{RMSE}_k^{-}$ are higher near the interfaces ($k \lesssim 30$ or $k \gtrsim 470$) compared to the bulk ($230 \lesssim k \lesssim 270$) for either models. This difference was also observed in the MSE values (Figure \ref{ml.train.test.loss} inset), and can be attributed to the greater complexity in the variations of the ionic structure near the interfaces due to changes in electrolyte attributes. 

The lower MSE values (Figure \ref{ml.train.test.loss}) for the ModNN model predictions for systems in the unseen test dataset show that the training of the surrogate using the modified loss function $\mathcal{L}$ leads to a more optimal selection of the weights and biases compared to training the surrogate using the standard loss function $L$.    
Further, the significantly smaller RMSE associated with the predictions made by the ModNN model (Figure \ref{fig.success}) demonstrate that the surrogate exhibits higher generalizability when trained with $\mathcal{L}$. 
In the following sections, we use the ML surrogate trained using the modified loss function $\mathcal{L}$ (ModNN model) to predict the ionic structure.

\subsection{Ionic Density Profiles}
We compare the cation and anion number density profiles predicted by the ML surrogate with the ground truth results obtained using MD simulations for input systems in the test dataset. 
Figure \ref{fig.predictions.density} shows the results for 4 input systems (I, II, III, IV) randomly selected from the test dataset. The systems are labeled by 5 input variables characterizing the ionic system (defined in Section \ref{sec:system.inputs.outputs}): $h$, $c$, $d_+$, $d_-$, $\sigma_{s}$. The 4 systems are: system I (4.8, 1.5, 0.415, 0.63, $-$0.02), system II (5, 2, 0.63, 0.63, $-$0.02), system III (4.4, 1.0, 0.415, 0.5225, $-$0.015), and system IV (4, 1.5, 0.5225, 0.415, $-$0.015). 

For each system, the ML-predicted ion density profiles are in excellent agreement with the ground truth results extracted using MD simulations.
A quantitative assessment of the accuracy of the predicted density profiles by the ML surrogate can be made by extracting the RMSE values for an input system $j$ for cation and anion density profiles: 
\begin{equation}\label{eq.rmsej}
\mathrm{RMSE}_j^{+ \, \mathrm{or} \, -} = \left(\frac{2}{P} \sum_{k=1}^{P/2} \left(y_j^k - \hat{y}_j^k \right)^2\right)^{1/2},
\end{equation}
where, $\hat{y}_{j}^{k}$ is the prediction of the cation or anion number density at the $k^{\mathrm{th}}$ bin, $y_{j}^{k}$ is the corresponding ground truth, and $P/2=502$ is the total number of predictions per density profile. 
$\mathrm{RMSE}_j^+$ values associated with the cation density profiles for systems I, II, III, and IV are 0.0007, 0.001, 0.0003, and 0.0004 respectively.
$\mathrm{RMSE}_j^-$ values associated with the anion density profiles for systems I, II, III, and IV are 0.0004, 0.0009, 0.0003, and 0.0005 respectively.    

\begin{figure*}
\centerline{\includegraphics[scale=0.5]{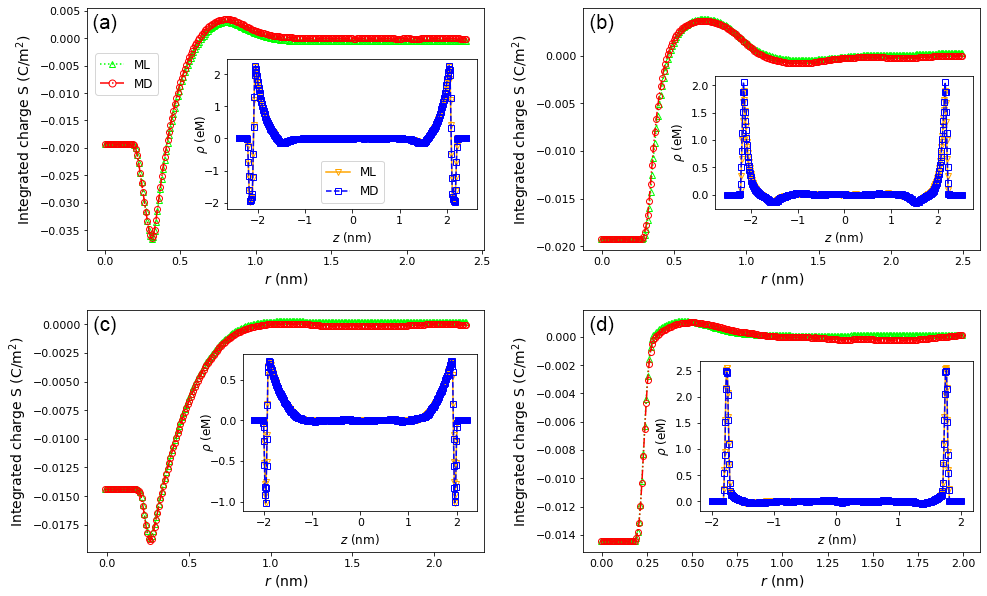}}
\caption
{\label{fig.predictions.int.charge}
Integrated charge $S$ produced by the ML surrogate (green up triangles) compared with MD simulation results (red circles) for electrolyte systems I (a), II (b), III (c), and IV (d). 
Inset shows surrogate predictions (orange down triangles) and simulation results (blue squares) for the associated net charge density $\rho$. 
See main text for the input system specifications.
}
\end{figure*}
    
It is useful to provide additional metrics to assess the accuracy of the surrogate predictions for the density profiles. Following our previous paper \cite{kadupitiya2020machine2}, we compute the success rate $A_j$ associated with the $j^{\textrm{th}}$ input system as:
\begin{equation}\label{eq.Ai}
A_{j} = \frac{1}{P}\sum_{k = 1}^{P}\Theta\left(\left|y_j^k - \hat{y}_j^k\right|, \epsilon_{j}^{k}\right),
\end{equation}
where $\hat{y}_{j}^{k}$ is the surrogate prediction of the ion density at the $k^{\mathrm{th}}$ bin, $y_{j}^{k}$ and $\epsilon_{j}^{k}$ are respectively the corresponding ground truth result and error obtained using MD simulations, and $\Theta$ is a step function defined as $\Theta(x,\epsilon) = 1$ for $x < \epsilon$; $\Theta(x,\epsilon) = 0$ for $x \ge \epsilon$. 
Note that the sum in Equation \ref{eq.Ai} is over the entire set of predictions $P = 1004$ made using the ML surrogate, which includes predictions for both cation and anion densities. 
The success rate $A_j$ can be thought of as the number of instances when the ML prediction of density values for the given ionic system $j$ were found to be within the error bounds (Equation \ref{eq:error}) produced by the associated MD simulation. For example, a success rate of $0.97$ implies that out of 1004 density predictions, 974 were within the associated uncertainties obtained from MD simulations.

The success rates $A_i$ for the prediction of density profiles for systems I, II, III, and IV are $0.975$, $0.93$, $0.98$, and $0.978$ respectively. These values are similar to the success rates computed in our previous paper for ionic systems characterized with uncharged surfaces and size-symmetric ions. 
Thus, the surrogate trained using the modified loss function yields high success rates despite the increase in the complexity of the ionic systems considered in this work. We note that the success rates using the BaseNN model (standard loss function $L$), are around $0.7$ for these 4 systems. 

In addition to the high accuracy of the surrogate, the inference time associated with ML predictions of density profiles ($\approx 0.1$ seconds) is a factor of $200,000\times$ smaller compared to the corresponding parallelized MD simulation runtime ($\approx 6$ hours). The combination of high accuracy and small inference times associated with the surrogate predictions provides rapid access to quantities derived using the density profiles and facilitates the implementation of complex analysis tasks. We will discuss these results next.

\subsection{Net Charge Density and Integrated Charge}
The excellent agreement between the cation and anion number densities generated via ML surrogate and MD simulations paves the way for using the surrogate to derive the net charge density $\rho$ (Equation \ref{eq:rho}) and the integrated charge $S$ (Equation \ref{eq:S}).
The integrated charge provides a way to quantify the effects of ion accumulation and depletion near the surface on the screening of the surface charge. 
We define $S$ using the left planar interface as the reference surface, and plot it as a function of $r = z + h/2$, where the latter denotes the distance from the left interface. Note that $S(r)$ is meaningful for $0 \le r \le h/2$ (or $-h/2 \le z \le 0$), i.e., for a distance $r$ less than half the confinement length. 

Figure \ref{fig.predictions.int.charge} shows the integrated charge $S(r)$ and the net charge density $\rho(z)$ (inset) computed by the ML surrogate for the same aforementioned four systems (I, II, III, and IV). The corresponding ground truth results obtained using MD simulations are also shown. 
For each system, the integrated charge and the net charge density computed by the ML surrogate are in excellent agreement with the ground truth results.

In a recent study \cite{anousheh2020ionic}, using the structural properties $\rho$ and $S$, we demonstrated the existence of two distinct regimes of screening behavior for different electrolyte systems as the electrolyte concentration $c$ increases from 0.1 M to 2.5 M.
While a broad range of systems was generated and explored by tuning $d_+$, $d_-$, $\sigma_s$, and $h$, the higher simulation costs limit the campaign to probe the material design space to hundreds of electrolyte systems. The availability of an accurate and fast ML surrogate enables a reliable access to variations in the ionic structure with continuous changes in the solution conditions for millions of electrolyte systems within minutes. This enables campaigns to probe the continuous input design space within the boundaries set by the initial simulation exploration used to design the surrogate. 

Figure \ref{fig.predictions.int.charge.diff.c} shows an example of a surrogate-led campaign generating the variation in the net charge density $\rho$ and the integrated charge $S$ with changing electrolyte concentration $c =$ 0.3, 0.5, 0.7, 0.9, 1.1, 1.3, 1.5, 1.7, and 1.9 M for an electrolyte system with $h = 5$ nm, $d_+ = 0.415$ nm, $d_- = 0.63$ nm, and $\sigma_{s} = -0.01$ C/m$^2$.
Results are shown for the left half of the confinement within 1.5 nm from the interface. Increasing $c$ leads to enhanced accumulation of cations and anions near the interface as evident by the rise in the magnitude of the peaks in $\rho$ near the interface. The pattern of the decay of $S$ to 0 as $r \to h/2$ also changes as $c$ is increased. For low $c$, $S(z)$ exhibits a monotonic decay with a characteristic decay length decreasing sharply with increasing $c$. However, for high $c$, $S(z)$ shows a non-monotonic oscillatory decay with a decay length exhibiting a small rise with increasing $c$. 
These findings produced by the ML surrogate track the behavior reported in our recent paper \cite{anousheh2020ionic}.

\subsection{Sensitivity Analysis}
\begin{figure}
\centerline{\includegraphics[scale=0.42]{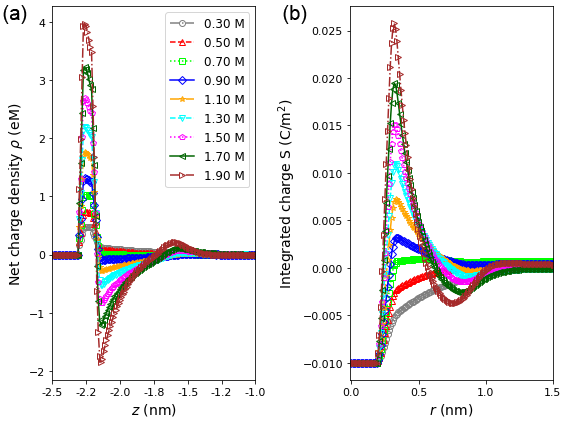}}
\caption
{\label{fig.predictions.int.charge.diff.c}
Evolution in the net charge density $\rho$ (a) and integrated charge $S$ (b) due to the changes in the concentration $c$ $\in [0.3, 2.0]$ M for an electrolyte system with other input variables held fixed at $h = 5$ nm, $d_+ = 0.415$ nm, $d_- = 0.63$ nm, and $\sigma_{s} = -0.01$ C/m$^2$ (see main text for the meaning of these symbols).
Increasing $c$ leads to more modulations in $\rho(z)$, and alters the pattern of decay of $S(r)$ to 0 as the distance $r$ from the interface is increased. 
}
\end{figure}

The rich variations observed in the net charge density and the integrated charge as a function of the location within the confinement (Figure \ref{fig.predictions.int.charge}), and the dramatic changes in these quantities with increasing concentration (Figure \ref{fig.predictions.int.charge.diff.c}) are linked to the complex organization of cations and anions within the confinement. 
For example, simulation snapshots extracted using the trajectory data indicate the emergence of structured layers of cations and anions near the interface when the concentration is increased to values $\gtrsim 1$ M \cite{anousheh2020ionic}. 
The high accuracy and small inference times associated with the ML surrogate predictions facilitate the implementation of complex analysis tasks for probing the observed modulations in the ionic structure and distilling the contributions of the attributes of cations and anions to the structural properties.

The surrogate can generate millions of cation and anion density profiles within minutes to yield smooth response surfaces describing variations of output structural properties to any changes in input electrolyte attributes. Leveraging this information, the surrogate can perform rapid sensitivity analysis and quantitatively assess the relative contributions of different ion attributes to the observed structural properties. Specifically, we demonstrate the use of the ML surrogate to perform sensitivity analysis and extract sensitivities of the response of the net charge density $\rho(z)$ to variations in the cation size $d_+$ and the anion size $d_-$ for different locations within the confinement.
    
The sensitivity analysis can be performed using many methods \cite{hunt2015puq,nossent2011sobol}. We use variance-based methods  \cite{im1993sensitivity,homma1996importance,nossent2011sobol}, wherein the variance of the output is decomposed into fractions that can be attributed to distinct input variables.
We present the process for computing the sensitivity of the output quantity $\rho$ for the case of two input variables: $d_+$ and $d_-$. The steps can be readily generalized to a large number of outputs and inputs. 
We begin by identifying the response surface $\mathcal{P}(D_+, D_-; z)$ connecting the distribution $\mathcal{P}$ associated with the output $\rho$ to the distributions $(D_+, D_-)$ associated with the input variables $(d_+, d_-)$, for each $z$. 
Generally, $(D_+, D_-)$ are randomly sampled from distributions with mean values specified by a fixed pair of $(d_+, d_-)$.

Let $E[\mathcal{P}]$ be the expectation value (mean) of $\mathcal{P}$. Let $V \equiv \mathrm{Var}[\mathcal{P}]$ be the associated variance. 
In the variance-based methods, we compute the contribution of varying only one input variable (e.g., $D_+$) to the overall output variance $V$, which can be described as an average over the variations in all the other input variables (e.g., $D_-$). 
We first define $E[\mathcal{P} | D^i_+] = 1/N_- \sum_{j=1}^{N_-} \mathcal{P}(D^i_+,D^j_-;z)$
as the average of the output net charge density $\rho$ for a fixed $D^i_+$ sampled from a distribution (e.g., normal). This average is computed by summing over $N_-$ number of samples drawn from a distribution associated with the input variable $D_-$.
The process is repeated for all $D^i_+$ values, and $E[\mathcal{P} | D^i_+]$ for $i = 1,2, \ldots, N_+$ are generated, where $N_+$ denotes the total number of $D_+$ samples.
The variance $V_+$ measuring the effect of varying only $D_+$ is the variance of this resulting output, $E[\mathcal{P} | D^i_+]$:
\begin{equation}
 V_+ = \frac{1}{N_+} \sum_{i=1}^{N_+} \left( E\left[\mathcal{P} | D^i_+\right] - E\left[\mathcal{P}\right] \right)^2.      
\end{equation}
The sensitivity $S_+$ quantifying the contribution of varying only $D_+$ to the overall output variance $V$ is 
\begin{equation}
\label{eq:splus}
S_+ = \frac{V_+}{V}.
\end{equation}
Following the same process and replacing $+$ with $-$, the sensitivity $S_-$ quantifying the contribution of varying only $D_-$ to the overall output variance $V$ is: 
\begin{equation}
\label{eq:sminus}
S_- = \frac{V_-}{V}.
\end{equation}
Note that $S_+ + S_- = 1$.

We now extract the sensitivities $S_+$ and $S_-$ for the net charge densities representative of the distinct regimes of screening behavior found under low $c$ and high $c$ conditions. We select two electrolyte systems, one at low $c=0.3$ M and another at high $c=1.9$ M; other input variables for both systems are fixed at $h=4.5$ nm and $\sigma_s = -0.015$ $\mathrm{C}/\mathrm{m}^2$. The mean values for the cation and anion diameter are taken to be $d_+ = 0.4$ nm and $d_- = 0.6$ nm respectively.
As the net charge density profiles $\rho(z)$ for these two systems involve the assessment of $\rho$ for a large set of $502$ $z$ values, one can extract a total of $502$ pairs of sensitivities $S_+$ and $S_-$. For the ease of illustration, we limit our analysis to 5 $z$ values within the left half of the confinement: (I) $z=-2.05$ nm, (II) $z=-1.95$ nm, (III) $z=-1.85$ nm, (IV) $z=-1.55$ nm, and (V) $z=-1.0$ nm. 
These values are chosen as representative of the regions within confinement where ionic size effects are expected to influence distinct features of the ionic structure. Let $r$ denote the distance from the left interface.
(I) and (II) represent the points of contact density of the cation ($r = d_+/2$) and anion ($r = d_-/2$) respectively, (III) represents a distance $r$ between $d_-/2$ and $d_-$, (IV) represents a distance $r$ slightly greater than  $d_-$, and (V) represents a point near the center of the confinement,  within the bulk region ($r=1.25$ nm) for either systems. 

\begin{figure*}
\centerline{\includegraphics[scale=0.385]{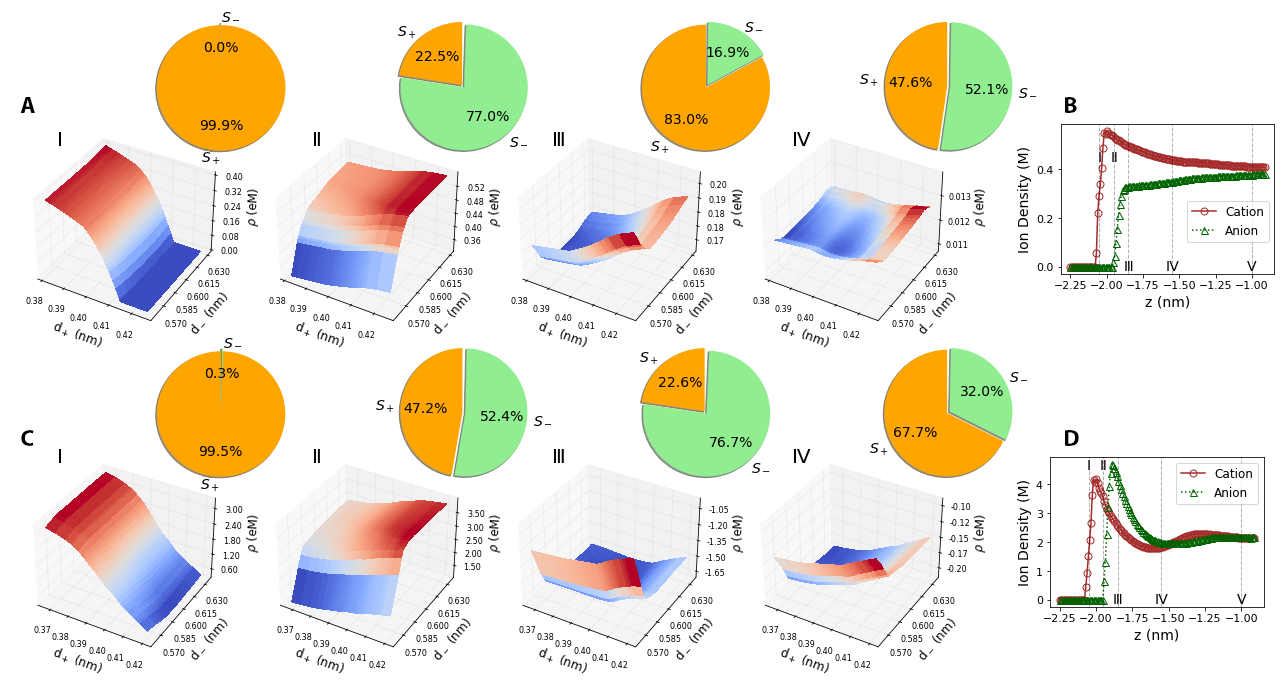}}
\caption
{\label{fig.predictions.response.sensitivity}
Sensitivity analysis results for the response of the net charge density $\rho$ to small variations in cation diameter $d_+$ (with mean value $0.4$ nm) and anion diameter $d_-$ (with mean value $0.6$ nm) for two electrolyte concentrations $c=0.3$ M (A and B) and $c = 1.9$ M (C and D). A and C show the response surfaces $\mathcal{P}$ (3D plots) and pie charts representing associated sensitivities $S_+$ to cation diameter (orange) and $S_-$ to anion diameter (light green) for the low and high $c$ systems respectively. 
B and D show the associated number density profiles $n_+(z)$ (brown circles) and $n_-(z)$ (green triangles) with $d_+ = 0.4$ nm, $d_- = 0.6$ nm for the low and high $c$ systems respectively.
Sensitivity analysis results are shown for four locations ($z$ values) within the confinement (columns I -- IV); the $z$ values are marked by black dashed lines in B and D.
}
\end{figure*}

We generate 1000 samples of cation sizes from the normal distribution $D_+ \sim \mathcal{N}(d_+,\,0.01^{2})$ nm, and 1000 samples of anion sizes from $D_- \sim \mathcal{N}(d_-,\,0.015^{2})$ nm, thus creating a collection of 1 million electrolyte systems. Note that $d_+ = 0.4$ nm and $d_- = 0.6$ nm. We then use the ML surrogate to predict the cation and anion number density profiles, and the associated net charge densities, for the entire set of 1 million samples. This process generates the response surfaces $\mathcal{P}(D_+, D_-; z)$ connecting the distribution $\mathcal{P}$ associated with the output $\rho$ to the distributions $(D_+, D_-)$ associated with input variables $(d_+, d_-)$ for the aforementioned five $z$ values. Leveraging the response surfaces, we extract the  sensitivities using Equations \ref{eq:splus} and \ref{eq:sminus}.
    
Figure \ref{fig.predictions.response.sensitivity} shows the response surfaces and associated sensitivities $S_+$ and $S_-$ for four $z$ values (I -- IV) for the low $c=0.3$ M (Figure \ref{fig.predictions.response.sensitivity}A) and high $c = 1.9$ M (Figure \ref{fig.predictions.response.sensitivity}C) electrolyte systems. We represent the sensitivities as percentages in the form of a pie chart next to the 3D plot of the response surface $\mathcal{P}$. The corresponding net charge density profiles $\rho(z)$ for the low and high $c$ systems are shown in Figure \ref{fig.predictions.response.sensitivity}B and Figure \ref{fig.predictions.response.sensitivity}D respectively.

At the point of contact density of cations (case I: $z = -2.05 \, \mathrm{nm}, \, r = d_+/2$), we find $S_+ > 99\%$ and $S_- \approx 0\%$ for both low and high $c$ systems. This indicates that the sensitivity of $\rho$ to changes in ion size is almost entirely dominated by variations in the cation size $d_+$ at the distance $r = d_+/2$ away from the interface. This finding is expected because anions, which on average are larger compared to cations in the system considered, are excluded from the neighborhood of $r = d_+/2$ (Figure \ref{fig.predictions.response.sensitivity}B and D), and therefore the steric effects influencing the ionic structure emerge entirely due to the cation size regardless of the electrolyte concentration. 
    
The contribution of the size of the anions to the sensitivity of $\rho$ becomes prominent near the point of contact density of anions (case II: $z = -1.95 \, \mathrm{nm}, \, r = d_-/2$). For $c=0.3$ M, we find $S_+ = 22.5\%, \, S_- = 77\%$, i.e., $\rho$ is much more sensitive to the variations in the anion size compared to those in the cation size. We attribute this to the depletion of anions resulting from the negatively-charged interface in the vicinity of $r=d_-/2$, as evident by the associated anion density profile (Figure \ref{fig.predictions.response.sensitivity}C). Small variations in the anion size alter the location of the anion contact density. Therefore anion density at $r= d_-/2$ changes with $d_-$ in a substantial way, making $\rho$ highly sensitive to variations in anion size compared to cation size. 
Results are different for $c=1.9$ M: $S_-$ is only marginally larger compared to $S_+$. The anion contact density is not as sensitive to the anion size because of the strong steric ion-ion repulsion that pushes anions near the interface regardless of their size and repulsion from the negatively-charged interface. 

At a point that represents a distance $r$ between $d_-/2$ and $d_-$ (case III: $z = -1.85$ nm), we observe a clear switch in the sensitivity profile between the low and high $c$ systems. For the low c system, $\rho(z)$ is much more sensitive to the changes in cation size compared to anion size ($S_+ = 83\%$; $S_- =16.9\%$). Here, the accumulation of cations is higher than anions (Figure \ref{fig.predictions.response.sensitivity}C). 
Changes in cation size lead to stronger variations in cation density compared to those in anion density resulting from changes in anion size.
The higher sensitivity of $\rho$ to changes in anion size in the high $c$ system can be attributed to the vicinity of $z = -1.85$ nm inhabiting the location of the anion peak density, which dominates the contribution to $\rho$ compared to the relatively depleted cation density (Figure \ref{fig.predictions.response.sensitivity}D). 
These findings suggest that the peak of anion density is highly sensitive to the anion size.
    
For distance $r$ slightly greater than anion diameter from the interface (case IV: $z=-1.55$ nm, $r = d_- + 0.1$ nm), $S_+ \approx S_-$ for the low $c$ system, indicating that the variations in cation and anion sizes contribute equally to the sensitivity of $\rho$. This can be attributed to the weak modulations in the ionic structure at distances greater than the anion diameter, which can be identified as the onset of the bulk region for the low $c$ system (Figure \ref{fig.predictions.response.sensitivity}C). Similar $S_+$ and $S_-$ values indicate that variations in cation and anion sizes contribute in equal amounts to changes in the bulk value of $\rho$.
For the same location within the confinement ($r = d_- + 0.1$ nm), $S_+ \approx 2S_-$ for the high $c$ system, indicating that the modulations in $\rho$ are still strong at this distance from the interface. At $r = d_- + 0.1$ nm, the high $c$ system has not reached the bulk region (also evident in Figure \ref{fig.predictions.response.sensitivity}D). The switch in the sensitivity profile compared to case III, where $d_-/2 < r < d_-$ and $S_- \gtrsim 3S_+$, originates from the presence of alternating layers of cations and anions near the interface. This structured region extends much farther into the center of the confinement compared to the low $c$ system. Only at distances greater than twice the anion diameter from the interface (case V: $z=-1$ nm, $r = 2d_- + 0.05$ nm), we find $S_+ \approx S_-$ for the high $c$ system, signaling the onset of the bulk region.

\section{Conclusion} \label{sec:conclusion}

Output quantities produced by molecular dynamics simulations of soft materials are generally associated with statistical uncertainties. We showed that these uncertainties can be utilized to informate the training of deep neural networks for designing machine learning surrogates aimed at predicting the relationship between input variables and simulation outputs. The approach was illustrated with the design of a surrogate for molecular dynamics simulations of confined electrolytes to predict the complex relationship between the input electrolyte attributes and the output ionic structure. 
We demonstrated that the prediction error for samples in the unseen test data can be significantly reduced by utilizing a modified loss function that leverages the uncertainties in the output ionic distributions during training. 

Using soft labels for the ground truth data facilitates a sampling mechanism that implicitly expands the dataset with more samples as the model undergoes training over many epochs, producing a surrogate that exhibits higher generalizability.
Comparing the improvement in accuracy resulting from the use of the modified loss function with results obtained using surrogates trained by explicitly expanding the training dataset size will be a subject of future work. 
Further, we will explore the extension of these ideas to develop surrogates for simulations of other soft materials exhibiting complex input-output relationships.

The surrogate predictions for the cation and anion number density profiles were found to be in excellent agreement with the ground truth results produced using molecular dynamics simulations. The high accuracy and small inference times associated with the surrogate predictions enabled rapid derivation of the net charge density and the integrated charge using the ion number densities, and facilitated the implementation of complex analysis tasks such as sensitivity analysis.
We demonstrated that the surrogate can rapidly extract sensitivities of the net charge density to cation and anion sizes, thus aiding in distilling the contributions of the attributes of cations and anions to the distinct ionic structure observed at low and high electrolyte concentrations.

\section*{ACKNOWLEDGMENTS}
This work is supported by the National Science
Foundation through Awards 1720625 and DMR-1753182, and by the Department of Energy through Award DE-SC0021418. Simulations were performed using the BigRed3 supercomputing system at Indiana University.



\end{document}